\documentclass[baaa]{baaa}
\usepackage[pdftex]{hyperref}
\usepackage{subfigure}
\usepackage{natbib}
\usepackage{helvet,soul}
\usepackage[font=small]{caption}
\contriblanguage{1} 

\contribtype{3}
\thematicarea{1}

\title{Clusters of Galaxies:}
\subtitle{Structure and Dynamics in the Last 8 Gyr}
\titlerunning{Dynamics of clusters of galaxies}
\author{A. Biviano\inst{1,2}}
\authorrunning{Biviano}
\contact{andrea.biviano@inaf.it}

\institute{INAF-Osservatorio Astronomico di Trieste, via Tiepolo 11, 34143 Trieste, Italy
\and
IFPU - Institute for Fundamental Physics of the Universe, via Beirut 2, 34014 Trieste, Italy}

\resumen{Los cúmulos de galaxias, al estar dominados por la materia oscura, han sido durante mucho tiempo y siguen siendo los objetivos cosmológicos ideales para estudiar la naturaleza de la materia oscura. Las informaciones sobre la naturaleza de la materia oscura provienen en particular de la determinación observacional de la distribución de la masa total y bariónica en cúmulos, en comparación con los resultados de simulaciones numéricas cosmológicas. Presento aquí resultados pasados y recientes sobre este tema de investigación.}

\abstract{Clusters of galaxies, being dark matter dominated, have long been and still are the ideal cosmological targets to study the nature of dark matter. Constraints on the nature of dark matter comes in particular from the observational determination of the distribution of total and baryonic mass in clusters, by comparison with results from cosmological numerical simulations. I present here past and recent results on this research topic.}


\begin{document}
\maketitle
\section{Introduction}
\label{s:intro}
\begin{figure}[!t]
  \centering
  \includegraphics[width=0.36\textwidth]{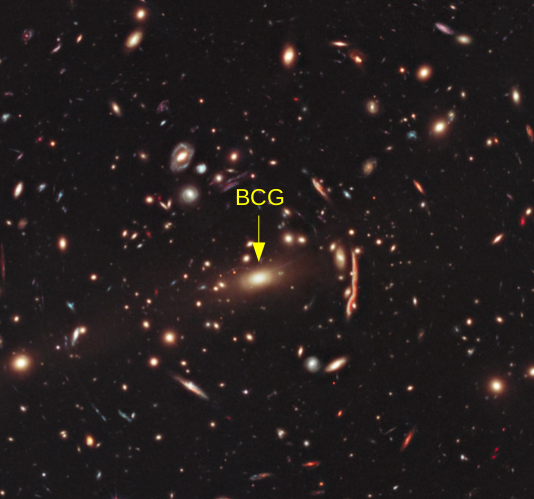}
  \caption{Image of the $z=0.44$ cluster MACS1206. The arrow indicates the Brightest Cluster Galaxy (BCG) that dominates the cluster center (credits: NASA, ESA, M. Postman.)}
  \label{f:bcg}
\end{figure}

Clusters of galaxies are the largest gravitationally bound systems in the universe, first catalogued by \citet{Abell58}. They were identified as overdensities of galaxies since the XVIII century and discovered as strong X-ray emitters in the 1970 \citep{Biviano00}\footnote{http://ned.ipac.caltech.edu/level5/Biviano2/frames.html}. After the discovery of X-ray emission from clusters of galaxies it became clear that most of the baryonic matter in clusters was not in galaxies but in the X-ray emitting diffuse, hot plasma. But most of the mass in clusters of galaxies is not in baryons, but in an unknown form of dark matter (DM hereafter). DM dominates the mass budget of clusters everywhere except near their centers, where many clusters
\citep[at least at redshift $z<1$,][]{vanderBurg+14} are dominated by the stellar mass of a massive ($\geq 10^{12} \, M_{\odot}$) central galaxy (the Brightest Cluster Galaxy, BCG hereafter, see Fig.~\ref{f:bcg}). Since their mass content is mostly in the form of DM, clusters of galaxies constitute an ideal laboratory to test for the nature of this yet undetected form of matter.

It was in clusters that the existence of DM was first postulated,
based on the observations of the relatively nearby and massive Coma cluster \citep{Zwicky33}. \citet{Zwicky33} applied the virial theorem to the observed spatial and velocity distribution of a few galaxies in the Coma cluster to estimate the cluster mass, and found it to be much larger than the light of the galaxies would suggest. More recently, observations of the so-called Bullet cluster in the optical and X-ray bands, allowed \citet{CGM04} to claim a direct detection of DM. Their claim was based on the observed displacement of the baryons (traced by the X-ray gas emissivity) and the gravitational potential (traced by weak lensing distortion of galaxies in the cluster background). Such a displacement cannot be explained by theories of modified gravity where there is no other mass component than the baryons.

\begin{figure}[!t]
  \centering
  \includegraphics[width=0.36\textwidth]{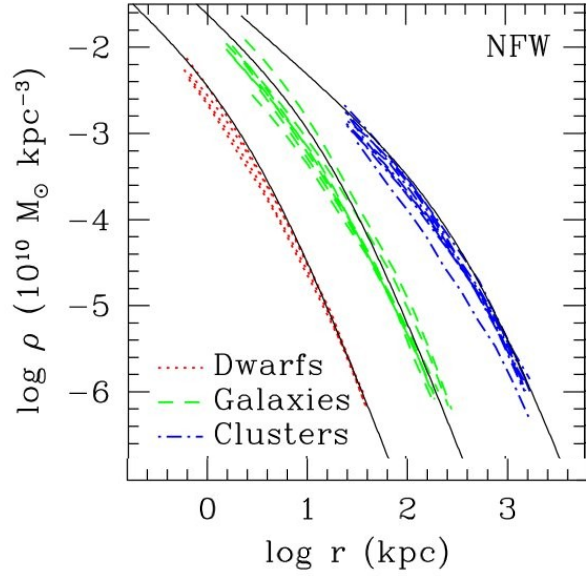}
  \caption{The mass density profiles of simulated halos of different masses are displayed together with their best-fit NFW model (solid black curves). From \citet{Navarro+04}.}
  \label{f:nfw}
\end{figure}

One way to explore the nature of DM is by comparison with predictions of cosmological numerical simulations feeded with a particular type of DM, the most popular choice being collision-less Cold DM. \citet{NFW96,NFW97} analysed the internal structure of simulated halos in Cold DM cosmological simulations and discovered a surprising similarity in the mass density profiles of halos of very different masses (see Fig.~\ref{f:nfw}). These profiles can be described by the simple law (NFW profile hereafter)
\begin{equation}
    \rho(r) \propto (1+r/r_{-2})^{-2} \, (r/r_{-2})^{-1},
\end{equation}
where $r_{-2}$ is the radius where the logarithmic derivative of the mass density profile  ${\rm d}\ln \rho / {\rm d}\ln r =-2$. It can also be written as $r_{-2}=r_{200}/c$, where $c$ is the concentration of the mass density profile, and $r_{200}$ is the radius where the mean mass density of the cluster is 200 times the critical density of the Universe. 

Deviation of the cluster mass profile from the NFW shape could indicate that DM differs from the Cold DM model that was used by \citet{NFW96} in their cosmological simulations. DM could be warm \citep{BOT01} or self-interacting \citep{SS00} rather than cold and/or collision-less. Deviations of the cluster mass profile from the NFW shape could also be due to baryonic effects, that were not included in the simulations of \citet{NFW96}. Several baryonic processes could affect the cluster mass profile; adiabatic contraction \citep{Blumenthal+86,Gnedin+04}, mass accretion \citep{Laporte+12,DK14,Schaller+15}, dynamical friction \citep{ElZant+01,ElZant+04}, and feedback from a giant active galactic nucleus (AGN hereafter) located in the BCG \citep{RFGA12,Peirani+17}.

\section{Measuring cluster mass profiles}
\label{s:jeans}
The determination of the mass distribution of clusters of galaxies can be obtained in several ways 
\citep[see, e.g.,][for a review]{Pratt+19}. The most popular are:
\begin{itemize}
    \item through the X-ray emission, or through the Sunyaev-Zel'dovich effect \citep{SZ69}, by assuming that the emitting plasma is in hydrostatic equilibrium in the gravitational potential;
    \item through gravitational lensing distortion of the shape of galaxies in the background of clusters;
    \item through the analysis of the projected phase-space distribution of cluster galaxies, used as tracers of the gravitational potential.
\end{itemize}
The first two methods work better with data taken from space-based observations, and are thus quite expensive. The third method requires extensive spectroscopy to measure the velocity distribution of cluster galaxies, and so it is generally based on ground-based observations. In this contribution I will present methods and results relative to the third method.

When galaxies are used as tracers of the gravitational
potential, the halo mass profile can be determined by application of
the spherical Jeans equation \cite{BT87},
\begin{equation}
  M(r) = - \frac{r \sigma_r^2(r)}{G} \left(\frac{\rm{d} \ln \nu}{\rm{d} \ln r}+\frac{\rm{d} \ln \sigma_r^2}{\rm{d} \ln r}+2 \beta(r)\right), \label{e:jeans}
\end{equation}
where $\nu(r)$ is the 3D galaxy number density profile, $\sigma_r(r)$
is the radial component of the 3D velocity dispersion profile, and
$\beta(r) \equiv 1-(\sigma_{\theta}^2+\sigma_{\phi}^2)/(2 \sigma_r^2)$, is
the profile of anisotropy of the galaxy velocity distribution,
$\sigma_{\theta}, \sigma_{\phi}$ representing the tangential components
of the 3D velocity dispersion profile, with the usual assumption
$\sigma_{\theta} \equiv \sigma_{\phi}$.

Equation~\ref{e:jeans} can be described as the equivalence between the gravitational pull of the cluster mass and the pressure resulting from the galaxy motions inside the cluster. Since these motions depend on the galaxy orbits, the function $\beta(r)$ enters the equation. Unfortunately we do not have direct observational access to 3D phase-space but only to projected galaxy positions and line-of-sight galaxy velocities. Given spherical symmetry it requires a simple integration to convert the projected galaxy number density profile into the 3D number density profile, $\nu(r)$, that enters eq.~\ref{e:jeans} \citep[the Abel integral, see][]{BT87}. On the other hand, to convert the observed, line-of-sight velocity dispersion $\sigma_{{\rm los}}$ to $\sigma_r$ requires knowledge of $\beta(r)$, since we need to know the relative strengths of the radial and tangential components of the velocity tensor to be able to deproject it from the line-of-sight observable. The solution for the Jeans equation is therefore generally considered to be degenerate between $M(r)$ and $\beta(r)$, the Mass-Anisotropy Degeneracy or Jeans' MADness. 

\begin{figure}[!t]
  \centering
  \includegraphics[width=0.46\textwidth]{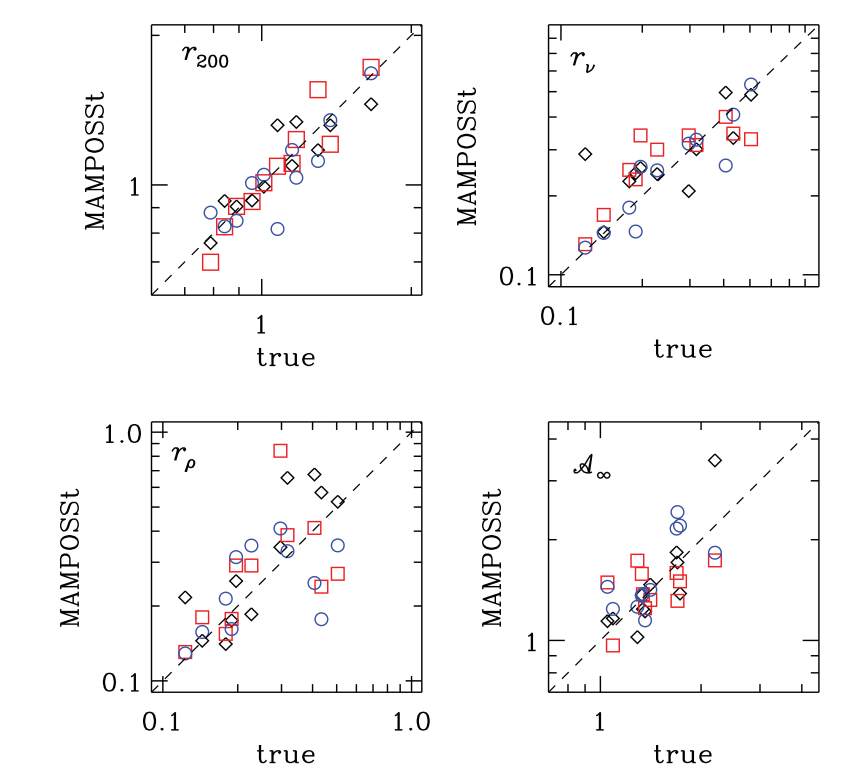}
  \caption{Correlation of MAMPOSSt-derived and true values of four jointly
fitted parameters ($r_{200}, r_{\rho}$ for M(r), $r_{\nu}$ for $\nu(r)$, ${\cal A}_{\infty}$ for $\beta(r)$) for several simulated halos with 500 tracers each. Each panel corresponds to a different
parameter, as labelled. Different symbols identify different projections. From \citet{MBB13}.}
  \label{f:mamposst}
\end{figure}

Following an early suggestion by \citet{Merritt87}, \citet{MBB13} developed an algorithm, named MAMPOSSt\footnote{{\em M}odelling {\em A}nisotropy and {\em M}ass {\em P}rofiles of {\em O}bserved {\em S}pherical {\em S}ys{\em t}ems}, to cure Jeans' MADness. The idea is that it is possible to reduce the Mass-Anisotropy Degeneracy by exploiting the full information available from the projected phase-space distribution, that is not restricted to the projected galaxies number density and line-of-sight velocity dispersion profiles. In practice, MAMPOSSt adopts parametrized models for $M(r), \nu(r),$ and $\beta(r)$ to estimate the probability of observing a cluster member at a given position in projected phase-space, i.e. at the observed projected distance from the cluster center, and with the observed cluster rest-frame velocity. By maximizing the products of the probabilities of all cluster members, MAMPOSSt finds the optimal parameters of the $M(r), \nu(r),$ and $\beta(r)$ models (see Fig.~\ref{f:mamposst}). 

Typically one considers only members located within the cluster virial radius (e.g. as estimated from a scaling relation) to ensure validity of the Jeans equation assumption of dynamical equilibrium. Also, the region dominated by the gravitational potential of the BCG (the central $\sim 30-50$ kpc) is typically excluded from the analysis, unless the models do explicitly account for the presence of the BCG and its gravitational potential in addition to that of the cluster. 

While MAMPOSSt is able to constrain the parameters of $M(r), \nu(r),$ and $\beta(r)$ simultaneously, in this contribution I focus on results for $M(r)$. Apart from the NFW model presented in Sect.~\ref{s:intro}, several other models have been proposed for the mass distribution of clusters of galaxies, and are generally used in the MAMPOSSt analysis. One is the generalization of the NFW profile (gNFW profile hereafter),
\begin{equation}
    \rho(r) \propto (1+r/r_{-2})^{\gamma-3} \, (r/r_{-2})^{-\gamma},
\end{equation}
where $\gamma$ is the value of the slope of the profile near the cluster center. Other profiles are characterized by different values of $\gamma$ and/or the asymptotic slope of the profile at large radii, $\gamma_{\infty}$. Relevant examples are the profiles of \citet[][Bur profile hereafter]{Burkert95} with $\gamma=0, \gamma_{\infty}=3$, of \citet[][Her profile hereafter]{Hernquist90} with $\gamma=1, \gamma_{\infty}=4$, and the softened isothermal sphere (SIS profile herafter) with $\gamma=0, \gamma_{\infty}=2$. Finally, the profile of \citet[][Ein profile hereafter]{Einasto65} is characterized by a slope that continuously changes with radius, $\gamma=2 \, (r_{-2}/r)^m$ with $m \simeq 5$ \citep[see, e.g.,][]{Biviano+13}.

\section{The data sets}
\label{data}
In this contribution I will present results from three data sets. 
\begin{itemize}
    \item WINGS \citep{Fasano+06}: 76 clusters at redshifts $z<0.2$, with spectroscopy from WYFFOS at the WHT and 2dF at the AAT  \citep{Cava+09}, with an average of 90 cluster members with positions and velocities per cluster \citep{Cava+17};
    \item CLASH-VLT \citep{Rosati+14,Caminha+17}: 12 clusters at $0.2<z<0.5$ with spectroscopy from VIMOS and MUSE at the VLT, totalling $\sim 8000$ cluster members with positions and velocities;
    \item GOGREEN \citep{Balogh+17}: 14 clusters at $0.9<z<1.5$, with spectroscopy from GMOS at the Gemini telescopes, totalling $\sim 500$ cluster members with positions and velocities, when including data from the GCLASS survey \citep{Wilson+09,Muzzin+12}.
\end{itemize}
\section{Results: the {\em total} mass profile}
\label{s:res-tot}
\subsection{Low-redshift clusters}
\label{ss:low}
\begin{figure}[!t]
  \centering
  \includegraphics[width=0.46\textwidth]{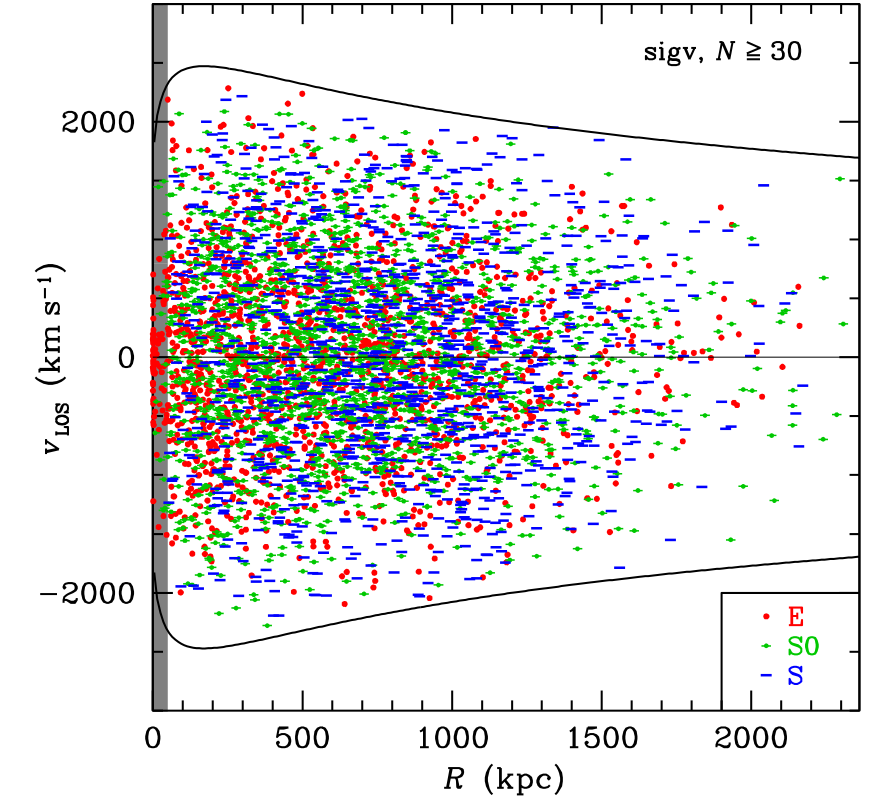}
  \caption{The projected phase-space distribution of cluster members in the stack of 54 WINGS clusters. Different symbols/colors represents galaxies of different morphological types. From \citet{Mamon+19}. Reproduced with permission from Astronomy \& Astrophysics, \copyright ESO.}
  \label{f:wingsRV}
\end{figure}
\begin{figure}[!t]
  \centering
  \includegraphics[width=0.46\textwidth]{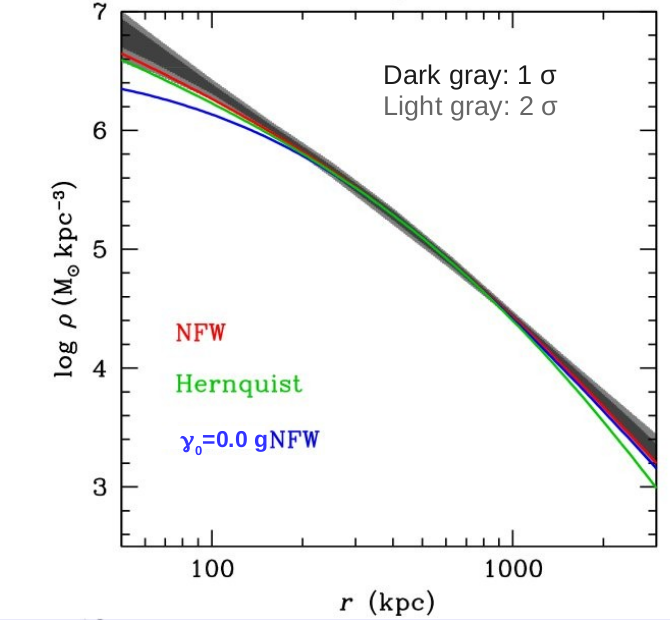}
  \caption{The total mass density profile of a stack of WINGS clusters as derived from MAMPOSSt adopting the gNFW model (grey shadings). Curves of different colors correspond to different models adopted in the MAMPOSSt analysis. From \citet{Mamon+19}. Reproduced with permission from Astronomy \& Astrophysics, \copyright ESO.}
  \label{f:wings}
\end{figure}
I describe here the recent results obtained by  \citet{Mamon+19} in their analysis of WINGS clusters. Their results are based on a stack of the subset of 54 WINGS clusters (see Fig.~\ref{f:wingsRV}) that have been found to be free of major internal substructures by the analysis of \citet{Cava+17} and that contain at least 30 member galaxies. Stacking of the different clusters is done by estimating a value of $r_{200}$ for each cluster, based on scaling relations of this parameter with either the line-of-sight velocity dispersion, or with the cluster X-ray temperature, or with the cluster richness (i.e. the number of cluster members within a certain range in rest-frame velocity and within a certain distance from the cluster center). Two different definitions were used for the cluster center, the location of the BCG and the peak of the X-ray emission. Results on $M(r)$ of the stack cluster are only mildly dependent on the choice of the scaling relation to determine $r_{200}$ and to the choice of the cluster centers. In the stack cluster, all cluster-centric distances are normalized by $r_{200}$, and all rest-frame velocities are normalized by $v_{200} \equiv 10 \, H(z) \, r_{200}$, where $H(z)$ is the Hubble constant at the mean redshift of the cluster.

The final stack pseudo-cluster contains more than 5000 members that are used to trace the gravitational potential. \citet{Mamon+19} considered many possible models in their MAMPOSSt analysis. They found that the highest likelihoods are obtained for gNFW models with steeper inner slope than NFW (see Fig.~\ref{f:wings}). 
However, there is no strong Bayesian evidence for a profile significantly different from NFW. 

\subsection{Medium-redshift clusters}
\label{ss:medium}
\begin{figure}[!t]
  \centering
  \includegraphics[width=0.46\textwidth]{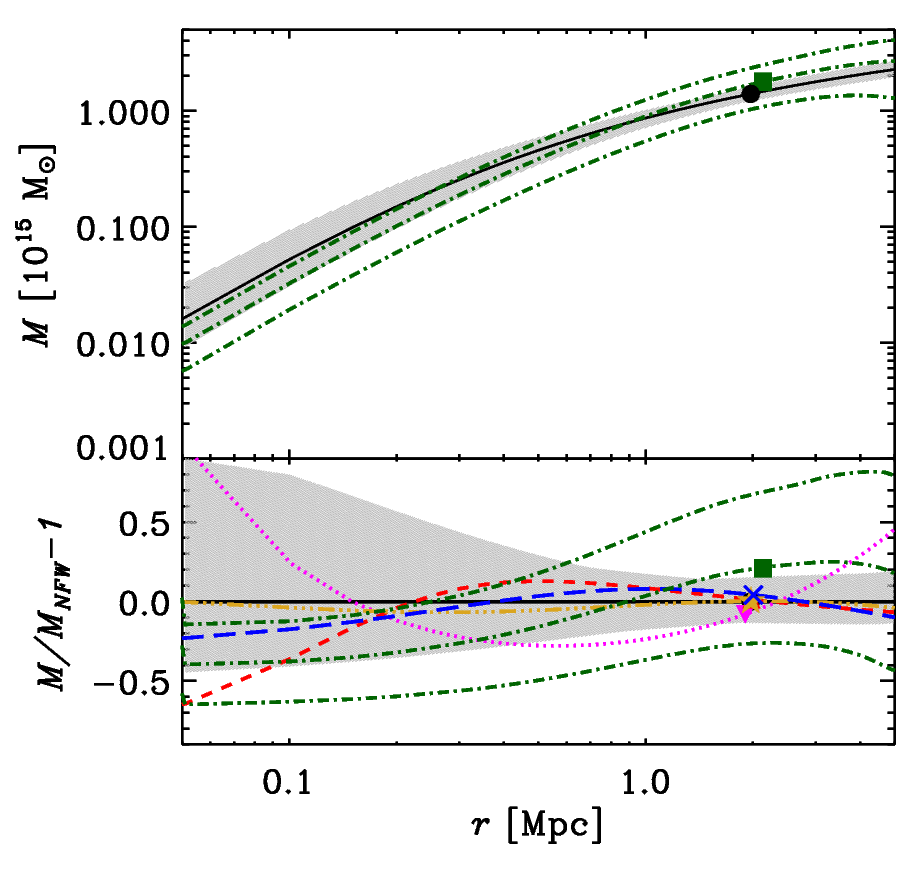}
  \caption{The total mass profile of the $z=0.44$ cluster MACS1206, as derived from MAMPOSSt. 
  {\em Top panel:} MAMPOSSt $M(r)$ best-fit NFW model (black curve) and 1 $\sigma$
    confidence region (grey shading). 
  {\em Bottom panel:} Fractional difference between different mass
    profiles and the MAMPOSSt best-fit to the NFW $M(r)$ (displayed in the top panel).  The
    Her, Ein, Bur, SIS models are represented by the blue
    long-dashed, gold triple-dot-dashed, and red short-dashed, and magenta curve,
    respectively. 
    The green lines in both panels represent results from the Caustic technique\citep{Diaferio99}, but they are not discussed in this contribution.
  From \citet{Biviano+13}. Reproduced with permission from Astronomy \& Astrophysics, \copyright ESO.}
  \label{f:clash}
\end{figure}
\begin{figure}[!t]
  \centering
  \includegraphics[width=0.46\textwidth]{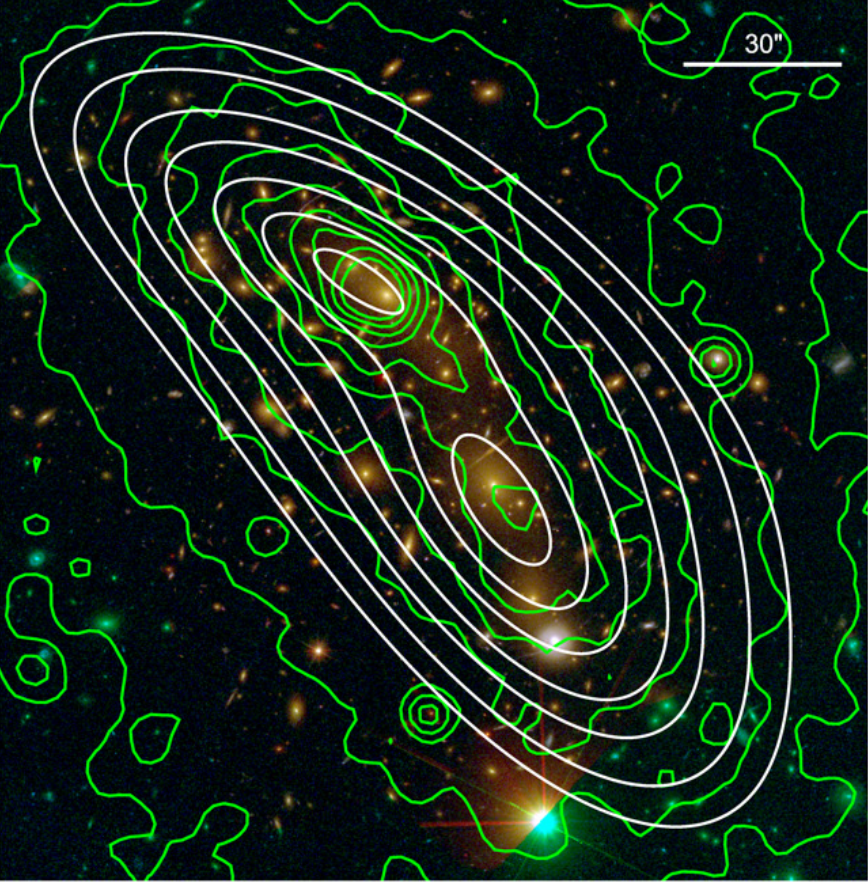}
  \caption{HST color image of the cluster MACS~J0416.1-2403 with extended mass halo from strong lensing modeling (white) contours and 0.5–2 keV X-ray emission (green) contours overlaid. 
  From \citet{Balestra+16}. Reproduced with permission from Astronomy \& Astrophysics, \copyright ESO.}
  \label{f:m416}
\end{figure}

At intermediate redshifts, the MAMPOSSt analysis has been applied so far to three of the CLASH-VLT clusters, all very massive, MACS~J1206.2-0847 at $z=0.44$ \citep{Biviano+13} with 600 cluster members, MACS~J0416.1-2403 at $z=0.40$ \citep{Balestra+16} with 800 cluster members, Abell~S1063 at $z=0.35$ (Sartoris et al., in prep.) with 1100 cluster members. We discuss Abell S1063 in Sect.~\ref{s:res-DM}. For the other two clusters, results from MAMPOSSt indicate that the mass profile of MACS~J1206.2-0847 is best fit by a NFW model (see Fig.~\ref{f:clash}) and cored models such as SIS and Bur are disfavored. On the other hand, the mass profile of MACS~J0416.1-2403 is best fit by a SIS model, that is a model with a central core. 

The different mass distribution between the two CLASH-VLT clusters might be attributed to their different dynamical status. MACS~J1206.2-0847 appears to be dynamically relaxed \citep{Girardi+15}, while MACS~J0416.1-2403 shows evidence of an ongoing merger of two massive sub-clusters near the cluster center \citep[see Fig.~\ref{f:m416}][]{Balestra+16}. Such a merger could temporarily affect the central mass distribution of the cluster, destroying its central cusp, as seen in other clusters \citep[e.g.,][]{Biviano+17b}.

\subsection{High-redshift clusters}
\label{ss:high}
\begin{figure}[!t]
  \centering
  \includegraphics[width=0.46\textwidth]{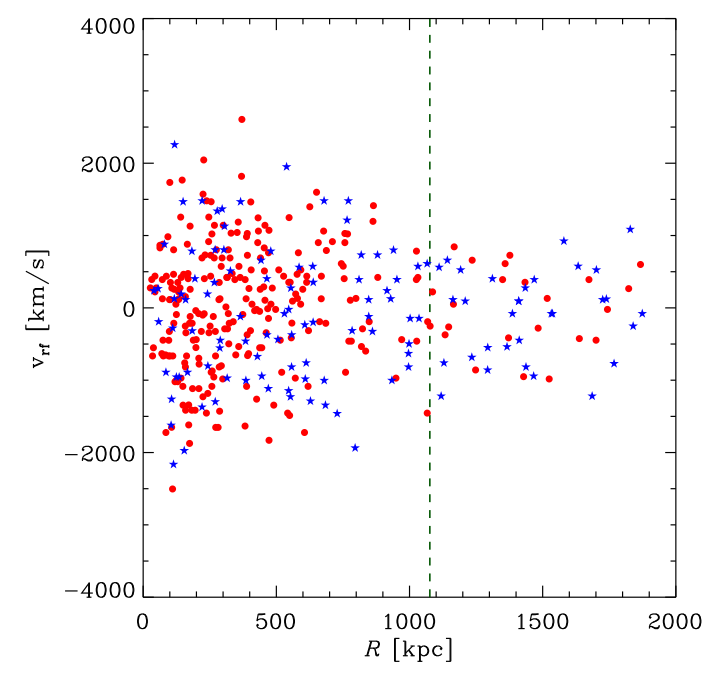}
  \caption{The projected phase-space distribution of cluster members in the stack of 10 GCLASS clusters. Blue/red points identify blue/red galaxies, resp. From \citet{Biviano+16}. Reproduced with permission from Astronomy \& Astrophysics, \copyright ESO.}
  \label{f:gclass}
\end{figure}
Using 10 clusters from the GCLASS survey at $0.87<z<1.34$, \citet{Biviano+16} constructed a stack cluster of 418 cluster members (see Fig.~\ref{f:gclass}), in a similar way as described in Sect.~\ref{ss:low}. Because of the clusters high-$z$, X-ray emission could not be measured, so neither the X-ray peak emission could be used to define the cluster centers, nor the X-ray temperature could be used to derive the cluster $r_{200}$. Centers were fixed at the cluster BCG positions, and $r_{200}$ values were derived from the scaling relation with line-of-sight velocity dispersions \citep{Munari+13}.

\citet{Biviano+16} analyzed this stack cluster with MAMPOSSt and found that its mass distribution is better fit by a Bur model than by an Ein, a Her, or a NFW model. While all models were found to be statistically acceptable in terms of their likelihoods, the fact that the Bur model is slightly favored by the data suggests that the slope of the central mass density profile is less steep than NFW, $\gamma<1$.

My ongoing MAMPOSSt analysis of the new GOGREEN data set, that includes the GCLASS sample, appears to confirm that $z \sim 1$ clusters are characterized by total mass density profiles that are flatter than NFW near the center. 

Taken together, results described in Sect.~\ref{ss:low}, \ref{ss:medium}, \ref{ss:high} suggest a steepening with time of the inner slope of the total mass density profile of clusters, a finding I discuss in Sect.~\ref{s:disc}.

\section{Results: the mass profile of the DM component}
\label{s:res-DM}
Results presented in the previous Section are for the {\em total} cluster mass profile. This includes not only the dominant mass component, DM, but also the baryonic components, namely the BCG stellar mass, the stellar and gas mass of all the other cluster galaxies, and the mass of the hot intra-cluster plasma 
\citep[see, e.g.,][]{BS06}. DM is dominant at all distances from the center except very near the center, where the stellar mass of the BCG can take over \citep{Sand+04}. If one wants to check the validity of the NFW model, it is the DM mass profile that must be determined, rather than the total mass profile. In fact, numerical simulations that include the baryonic components and not only DM, have shown that the inner slope of the total mass density profile can be steeper than NFW in nearby clusters \citep{Schaller+15}, in qualitative agreement with our findings (see Sect.~\ref{ss:low}).

As a matter of fact, the total mass distribution of the low-$z$ WINGS clusters (described in Sect.~\ref{ss:low}) is almost equally well fit by a gNFW model with $\gamma>1$ or by a combination of two NFW models, one describing the cluster and the other the BCG mass distributions \citep{Mamon+19}. This result suggests (but does not prove) that it is the BCG baryonic mass that makes the inner slope of the total mass density profile $\gamma>1$, at least at low-$z$. At higher $z$ the mass contribution by the BCG might be less important if the BCG formation time is delayed with respect to its cluster host as indicated by some observations \citep{DeMaio+19}.

\begin{figure}[!t]
  \centering
  \includegraphics[width=0.46\textwidth]{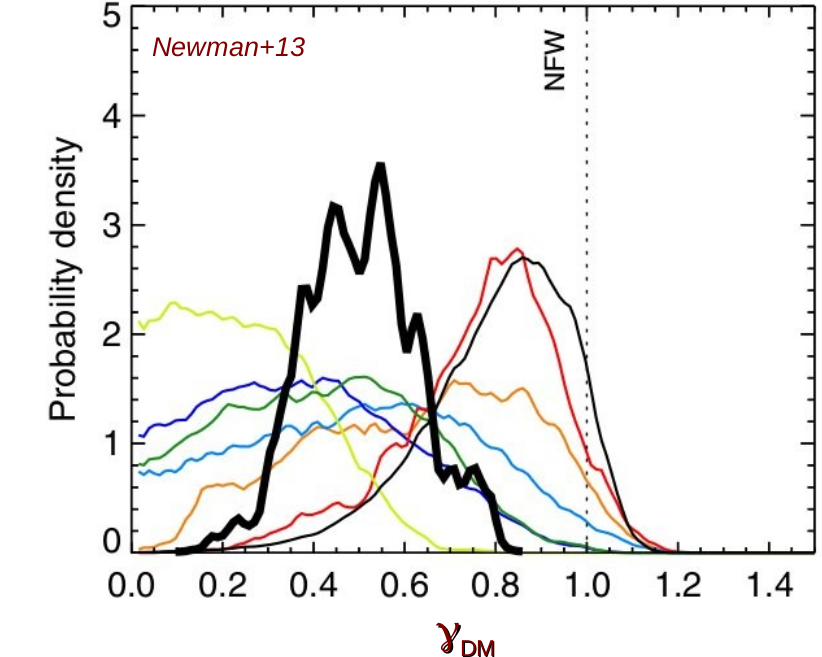}
  \caption{The probability distribution of the inner slope of the DM mass density profile $\gamma_{DM}$ for several medium-$z$ clusters. The combined probability distribution is shown by a thick black curve. From \citet{Newman+13}.
  $\copyright$ AAS. Reproduced with permission.}
  \label{f:newman}
\end{figure}

Disentangling the baryonic component from the total mass to determine the DM mass distribution requires an accurate and precise characterization of the different baryonic components, in particular of the BCG. In fact, it is the study of the inner cluster regions that is most important to discriminate among different kinds of DM. In a series of papers, \citet{Sand+04,Sand+08} and \citet{Newman+13} have used the combined information from the internal kinematics of the BCG and gravitational lensing, to determine the inner slope $\gamma_{DM}$ of the DM mass density profile. They found $\gamma_{DM}=0.5$, averaged over 7 clusters, with a random error of $\pm 0.1$ and a slightly larger systematic error (see Fig.~\ref{f:newman}). Such a value is inconsistent with NFW.

This claimed discrepancy between the observed value of $\gamma_{DM}$ and the expected value from DM-only simulations, has been addressed in the paper by \citet{He+19}. Using halos from the Cluster-EAGLE hydrodynamic simulations they confirmed the approximately NFW inner slope of the DM mass distribution of cluster-size halos. At the same time, they tested and validated the observational procedure of \citet{Newman+13} to determine $\gamma_{DM}$. The discrepancy between simulated and observed values of $\gamma_{DM}$ can only be solved, according to \citet{He+19}, if the scale radii $r_{-2}$ of the clusters analyzed by \citet{Sand+04,Sand+08} and \citet{Newman+13} were over-estimated (by 50\%), because an over-estimation of $r_{-2}$ propagates into an under-estimation of $\gamma_{DM}$ due to significant covariance between these two parameters.

Within the CLASH-VLT team we are using new and archival MUSE data for the BCG of some CLASH-VLT clusters to determine their values of $\gamma_{DM}$. Our approach is similar but somewhat different from the one adopted by \citet{Sand+04,Sand+08} and \citet{Newman+13}. We are combining the information coming from the kinematics and surface brightness of the BCG, observed with MUSE, with the information coming from the projected phase-space distribution of cluster galaxies, observed with VIMOS. The likelihood coming from MAMPOSSt is combined with the likelihood coming from the fit to the binned line-of-sight velocity dispersion profile of the BCG.

Preliminary results for two CLASH-VLT clusters, Abell~S1063 (Sartoris et al., in prep.) and MACS~J1206.2-0847 (Biviano et al., in prep.) indicate $\gamma_{DM}$ values in better agreement with the theoretical predictions of \citet{He+19} than the previous determinations by \citet{Sand+04,Sand+08} and \citet{Newman+13}.

\section{Discussion}
\label{s:disc}
The analysis of the projected phase-space of cluster galaxies shows that the cluster total mass density profile is almost invariant and not very different from the NFW model since $z \sim 1$, that is, over the last 8 Gyr of cosmic history. This leaves $\lesssim 2$ Gyr time for clusters to achieve their final stage of dynamical equilibrium since their formation \citep[the most distant formed cluster known is at $z=1.8$,][]{Andreon+14}. The cluster dynamical evolution after their formation must therefore be rapid, and this can be achieved via the violent \citep{LyndenBell67} or discreteness-driven relaxation \citep{BdSV19}.

Despite the fact that the total cluster mass is DM dominated, there are different theoretical expectations for the central logarithmic slope of the {\em total} and the {\em DM} mass density profile. A slope slightly larger than 1 is expected for the total mass density profile of massive clusters of galaxies \citep{Schaller+15}. A value $\gamma>1$ is indeed found for the best-fit mass model of nearby clusters from WINGS \citep{Mamon+19}. However, there is a tendency for total mass density profiles to be flatter near the cluster center at higher redshifts. Possible explanations for this flattening include the following;
\begin{itemize}
    \item mis-centering; we do not know the X-ray peak emission of distant clusters, so we must rely on the BCG position only, and at times, BCG could be offset from the bottom of the gravitational potential well.
    \item Dynamical disturbance; ongoing merger activity of two sub-clusters can at least temporarily remove a central cusp in the mass distribution. Since high-mass clusters form relatively recently \citep[e.g.][]{LC11}, clusters at high-$z$ are more likely to be observed in a disturbed phase of ongoing merging than clusters at low-$z$.
    \item AGN feedback \citep{RFGA12,Peirani+17}; the energy input of the BCG AGN is able to reduce the concentration of the intra-cluster plasma near the cluster center. By consequence of gravitational interaction, also the DM distribution near the center is similarly affected. Such an AGN feedback may be more important at a stage when the BCG is still forming. 
    \item Decreasing relative importance of the BCG baryonic component with increasing $z$. There is observational evidence that BCG may still be forming in high-$z$ clusters \citep{Miley+06,vanderBurg+15} and its mass assembly is delayed relative to that of its host cluster \citep{DeMaio+19}.
\end{itemize}

Determining the DM-only, rather than the total, mass density profile from observations, is less straightforward. One needs to separately account for the different baryonic mass components, the BCG stellar mass, the intra-cluster plasma, and the gas and stellar mass in the other cluster galaxies. Since it is the central behavior of the DM density profile that can distinguish between different types of DM, it is the BCG baryonic contribution the most important one in this kind of analysis. So far, strong constraints have only come from one research group, and they indicate $\gamma_{DM} < 1$ \citep[][and references therein]{Newman+13}. \citet{He+19} have used the Cluster-EAGLE simulations to confirm the early predictions by \citet{NFW96}, $\gamma_{DM} \approx 1$. \citet{He+19} have also shown that the observational procedure adopted by \citet{Newman+13} should lead to the correct values. \citet{He+19} mentioned the possibility of solving the discrepancy if the estimates of the cluster scale radii by \citep{Newman+13} are biased low, but they did not prove this to be the case.

So, while there is agreement between theory and observations on the total mass distribution of clusters, this is not the case for the DM mass distribution. This could have a strong impact on our understanding of the nature of DM. It is therefore very welcome that another group, the CLASH-VLT team, is addressing this issue, by using new MUSE data for the central cluster region coupled with VIMOS data for cluster galaxies out to large distances from the center, and an extension of the MAMPOSSt technique to include the information on the kinematics of the BCG. Preliminary results for two clusters indicate values for $\gamma_{DM}$ in agreement with the theoretical expectation from \citet{He+19}.

\section{Conclusions}
\label{s:conc}
Clusters of galaxies, being DM-dominated systems, can be very powerful probes to ascertain the nature of DM. We can determine their mass distribution from the central kpc to several Mpc from their center, by combining the information coming from the projected phase-space distribution of cluster galaxies, and from the velocity dispersion and surface brightness profile of the BCG. The current observational evidence appears to give conflicting results when the {\em total} or the {\em DM}-only mass distributions are considered. New MUSE observations of the central regions of CLASH-VLT clusters will improve our constrain on the DM inner mass profile. Additional constraints might come from existing observations of nearby clusters \citep{Rines+13,Loubser+18}. In parallel, it is important to use cosmological hydrodynamic numerical simulations such as DIANOGA \citep{Rasia+15,Biffi+16,Ragone+18} to understand the effects of baryons on the DM cluster mass profiles at different redshifts, and at different stages of the cluster evolution, and to validate the observational procedure against possible systematic biases, for instance those that could be induced by projection effects.

\begin{acknowledgement}
I wish to thank all my collaborators in the WINGS, CLASH-VLT, and GOGREEN teams, and in particular M. Balogh, A. Cava, G. Mamon, A. Mercurio, A. Moretti, B. Poggianti, P. Rosati, and B. Sartoris.
\end{acknowledgement}


\bibliographystyle{baaa}
\small
\bibliography{biblio}
 
\end{document}